\documentstyle[12pt,amssymb]{article}

\newcommand{\return}{\cr\noalign{\vspace{5pt}}}

\begin{document}

\begin{titlepage}

\def\thefootnote{\ast}

\begin{flushright}
Preprint~ISC~95--1 \\
February 1995
\end{flushright}
\vspace{20pt}

\begin{center}
{\Large\bf
On higher-derivative dilatonic gravity \\[10pt]
in two dimensions}\\[40pt]
S.~N{\sc aftulin} \\[5pt]
{\it  Institute for Single Crystals, 310141 Kharkov, Ukraine} \\[20pt]
S.D.~O{\sc dintsov}
\footnote{Electronic address: \   sergei@ecm.ub.es} \\[5pt]
{\it Department ECM, Faculty of Science, Barcelona University, \\[-2pt]
Diagonal 647, 08028 Barcelona, Spain}
\end{center}
\bigskip

\begin{abstract}
We discuss lowering the order of the two-dimensional scalar-tensor $R^2$
quantum gravity, by mapping the most general version of the model to a
multi-dilaton gravity, which is essentially the sigma-model coupled with the
Jackiw-Teitelboim-like
                  gravity. In the continuation of our previous research, we
calculate the divergent part of the one-loop effective action in a
$2D$ scalar-tensor (dilatonic) gravity with the $R^2$-term, which belongs to
a specific degenerate case and cannot be obtained from the general
expression.  The corresponding finiteness conditions are found.
\end{abstract}

\end{titlepage}

\def\thefootnote{\arabic{footnote}}
\setcounter{footnote}{0}

\section{Introduction}

Two-dimensional models of gravity are widely recognized to provide a much
deeper understanding of the quantum gravitational effects, even beyond the
scope of the perturbation theory (for a review, see \cite{TASI_lectures}).
Recently there has been a lot of activity in studying the dynamical structure
of various models of gravity and their connection to the strings. However,
most efforts were undertaken within the conventional (second order in
derivatives) models. The dynamical structure of a higher-derivative theory
is greatly complicated, even in two space-time dimensions. The most general
action of the two-dimensional ($2D$) fourth order scalar-tensor gravity
introduced in Ref.\cite{our_PLB94}
contains six different terms with four derivatives:
\begin{eqnarray}
S=\int\!d^2x\,\sqrt{-g}\,\Bigl\{Z_1(\Phi)(\partial^\mu\Phi\partial_\mu\Phi)^2
         +Z_2(\Phi)(\square\Phi)(\partial^\mu\Phi\partial_\mu\Phi)+Z_3(\Phi)
         (\square\Phi)^2\ \cr
   +Z_4(\Phi)R\partial^\mu\Phi\partial_\mu\Phi+Z_5(\Phi)R\square\Phi+Z_6
   (\Phi)R^2+\ldots \Bigr\} \ ,
\label{general_model}
\end{eqnarray}
where we have suppressed the lower order terms. In general, the potential
functions $\,Z_1,\ldots,Z_6\,$ are smooth but otherwise arbitrary.

There are several motivations to consider fourth-order theories in two
dimensions. First and foremost, they serve as a convenient playground for
studying much more intricate four-dimensional models. It is well known that
the Einstein gravity is not renormalizable in four dimensions so one often
has to resort to the $R^2$-theory (see, e.g., a book
\cite{Buchbinder_etal:book}
for a comprehensive review); the latter {\em is\/} renormalizable, and
asymptotically free, but suffers from non-unitarity already at the tree level
\cite{Stelle}.
In two dimensions, even if the theory does not look unitary, string-inspired
arguments may be developed \cite{ghost_decoupling_in_HD-gravity}
to show that the negative-norm states decouple at the renormalization group
(RG) fixed point.Thus, the study of 2D higher derivatives gravities
may provide some idea to solve the unitarity problem of
such theories in 4D.We show that in formulation with additional
scalars there are no problems with massive ghosts as theory
maybe mapped to equivalent low derivative model.
Second, the simulation data for 2D higher derivative gravity
models are getting avaliable \cite{boul}.Hence, such theories
maybe good testing models of quantum gravity
where comparison with numerical datas maybe done.Third,
2D higher derivative terms maybe relevant for problem
of the branched polymer \cite{string_susceptibility}.Fourth,
due to absence of 2D Einstein term there are following possibilities
to introduce the kinetic term for graviton-in non-local
Polyakov form, in the form of coupling of Einstein term with
dilaton or with help of higher derivative term.
                  Note also  that in
                        the effective theory of the string with the first
massive level taken into account %
\footnote{See, e.g., \cite{Buchbinder_etal:preprint}
for a recent discussion and further references.}
there may also appear higher derivative terms.

Some speculative connection between two- and four-dimensional ($4D$) theories
of gravitation is offered by a dimensional continuation
\cite{E-expansion_in_gravity}:
one starts with the $2+\epsilon$-dimensional version of quantum gravity and
performs the $\epsilon$-expansion, in the end the limit $\epsilon\to2$ may
be taken in order to get some insight into the non-perturbative coupling
effects.

Logically, there are two possible points of departure in this construction,
namely, the gravity in $4-\epsilon$ or in $2+\epsilon$ space-time dimensions.
It is reasonable to expect that the properties of both versions have some
overlap at intermediate values of $\,\epsilon\,$.

Unfortunately, such a matching procedure is not immediate since the origin
of the quantum gravity in $2D$ may be very different from that in $4D$: the
former is usually considered as the induced quantum gravity (which takes its
roots in the string theory) while the latter is not. Hence, if the
continuation in $\epsilon$ is to be addressed properly one has to take
special care about the $4D$ theory to be matched.

Some time ago, an infrared quantum $4D$ gravity was introduced in
Ref.\cite{infrared_quantum_gravity}:
The trace anomaly of matter fields in the curved space-time may be integrated
to yield an effective action for the conformal factor of the metric,
analogous to the Polyakov action \cite{dilaton_gravity_from_anomaly}
for $D=2$; the resulting expression describes a fourth-order scalar-tensor
theory which seems to be a more suitable candidate for matching across two
dimensions ($\epsilon\to2$).

There are apparently two ways of descending the $4D$ induced quantum gravity
to $D=2$: the first one is in the parameter space (i.e., by the continuation
in $\epsilon$) and the second is by some dimensional reduction. From the
first approach one probably concludes that the appropriate $2D$ theory
possesses the second order scalar-tensor Lagrangian, while from the second
approach it follows that it is of the fourth order. For these two
descriptions to be consistent, one has at least to show that a fourth-order
action (\ref{general_model}) may be re-written as some second-order one. The
latter problem is addressed in section~2: making use of a number of auxiliary
scalars, $\,\Psi_j\,$, Eq.(\ref{general_model}) is converted into the
string-inspired form:
\begin{equation}
S=\int\!d^2x\,\sqrt{-g}\,\left[{1\over2}{\cal G}_{ab}(\Phi)\partial^\mu\Phi_a
         \partial_\mu\Phi_b+{\cal B}(\Phi)R+{\cal U}(\Phi) \right] \ ,
\label{sigma_model}
\end{equation}
where we have denoted $\Phi_a\equiv\left\{\Phi;\,\Psi_j\right\}$. The absence
of (perturbative) spin-two states in the $2D$ quantum gravity ensures that
the tensor auxiliary fields need not be introduced.

At the risk of belaboring the obvious, let us emphasize the following. The
vacuum target manifold of the ``string'' (\ref{sigma_model}) is essentially
curved because of the non-trivial embedding constraints. There is not enough
freedom of reparametrization to nullify both non-tachyonic beta-functions,
$\,\beta_{B}\,$ and $\,\beta_{G}\,$ (in accordance with the observation that
the leading terms in the action have dimensions greater than $D$), thus
severe restrictions are to be imposed on the potential functions if the
conformal invariance is assumed to hold. This makes a crucial difference
with a conventional dilaton gravity where the target-space metric,
$\,{\cal G}_{ab}\,$, is flat and the dilaton, $\,{\cal B}\,$, is linear in
the string co-ordinates so that the possible counterterms are solely due to
the tachyon, $\,{\cal U}\,$, \cite{dilaton_gravity_as_string}.

The one-loop counterterms in the most general version of the fourth-order
$2D$ gravity were reported in our earlier work \cite{our_PLB94};
a discussion of some subtle points connected with the quantum gauge fixing
may be also found there. Section~3 below contains the evaluation of
divergences in a special (degenerate) class \cite{degenerate_R2-gravity}
of this theory, which cannot be obtained from the general formulas of
Ref.\cite{our_PLB94}.
We also find the family of the dilaton potential functions that guarantee
one-loop finiteness. The purely metric $R^2$-gravity in two dimensions
\cite{Yoneya},
coupled to $N$ conformal matter scalars,
\begin{equation}
S=\int\!d^2x\,\sqrt{-g}\,\Bigl[\omega R^2+CR+\Lambda+{1\over2}\partial^\mu
                                      \chi_j\partial_\mu\chi_j\Bigr] \ ,
\label{Yoneya_model}
\end{equation}
belongs to this set: it displays the critical behavior for arbitrary
$\omega$, $C$, and $\Lambda\,$ \cite{our_IJMPA}.
The divergences are proportional to the Euler characteristic,
$\,\smallint\!d^2x\sqrt{g}R\,$, of the space-time, with the effective central
charge $\,c=12-N\,$.  There is a hope that the value of $\,c\,$ persists to
higher orders in the loop expansion.Note that discussion of
generalization of above model for gravitational-torsionful
background maybe found in ref.\cite{vol}.

\section{Multi-dilaton gravity from an auxiliary field construction}

The first pattern of lowering the order in the higher-derivative gravity
was due to Yoneya \cite{Yoneya},
who showed that the purely metric action (\ref{Yoneya_model}) could be recast
as essentially the JT-like gravity           \cite{JT-model},
by the use of an auxiliary field, $\Psi$, representation:
\begin{equation}
\bigl[\,\ldots\,\bigr]^2=-\Psi^2-2\Psi\bigl[\,\ldots\,\bigr] \ .
\label{Yoneya_decomposition}
\end{equation}
It is important to realize that, as opposed to the ``customary'' auxiliary
fields, $\,\Psi\,$ acquires the kinetic term already at the tree level, due
to mixing with the conformal mode via the contact term $\,\sqrt{g}R\Psi\,$,
\cite{JT-model}.
Diagonalizing this, mixed, kinetic matrix one finds that the signs of the
eigenmodes are opposite so that the model has zero dynamical degrees of
freedom on shell. This agrees with the result of an explicit canonical
counting of the degrees of freedom in (\ref{Yoneya_model}), \cite{Yoneya}.

Consequently, counting of the degrees of freedom can give us a hint of how
many auxiliary scalar fields, $\,\Psi_j\,$, have to be introduced. In our
case, (\ref{general_model}), the straightforward counting gives two degrees
of freedom since the actual difference is one dilaton field, $\Phi$, entering
at the fourth order. (Note that in the fourth-order theory, the number of
degrees of freedom is effectively doubled \cite{Stelle}
as compared to the ordinary one.) So let us firstly try two auxiliary fields,
by Yoneya's decomposition (\ref{Yoneya_decomposition}):
$$
S=\int\!d^2x\,\sqrt{-g}\,\Bigl\{\lambda_1\bigl[R+\lambda_2(\partial\Phi)^2
                        +\lambda_3\square\Phi\bigr]^2+\lambda_4\bigl[\square
\Phi+\lambda_5(\partial\Phi)^2+\lambda_6R\bigr]^2+\ldots\Bigr\} \ ,
$$
(Note that similar suggestion to use two scalars for lowering the
order in this model has been given recently in ref.\cite{sha}).
There are six unknown functions, $\lambda_1(\Phi),\ldots,\lambda_6(\Phi)$,
to be expressed in terms of six dilatonic ``potentials'',
$Z_1(\Phi),\ldots,Z_6(\Phi)$. It is easily shown that no solution to this
algebraic problem exists: the simplest way to proceed is to put $\,Z_5=0\,$
from the outset, since an arbitrary $Z_5(\Phi)$ can be made zero all the
same, by an appropriate conformal metric rescaling (the other $Z$'s also
change under such a transformation). In the sigma-model language of
Ref.\cite{our_PLB94}
this means that $Z_5$ defines a Stueckelberg field,
\cite{Buchbinder_etal:preprint}.
To conclude, two auxiliary fields are not sufficient to lower the order of
the derivatives in (\ref{general_model}).

This fact is also understood from a trivial observation that the action
(\ref{general_model}) makes up a quadratic form on ``vectors''
$\,V\equiv\lambda_1R+\lambda_2(\partial\Phi)^2+\lambda_3\square\Phi\,$, so
one expects to have three independent linear combinations, and
correspondingly three auxiliary fields. The only possible reconciliation with
the direct counting of the degrees of freedom is that one such a field does
not admit a conventional tree-level propagator.

To show this in more detail let us start with the conformal metric
rescalings $\,g_{\mu\nu}\to g_{\mu\nu}\exp(\sigma)\,$. Due to the property
$\,\sqrt{g}R\to\sqrt{g}(R-\square\sigma)\,$ different $V$'s mix up under such
a transformation. So it is preferable to organize ``completing the square''
(\ref{Yoneya_decomposition}) in such a fashion that the respective
coefficients do not change, modulo the overall multiplication by
$\,\exp(-\sigma)\,$. Fortunately, two such functions are known: these are the
principal minors in the fourth-order kinetic matrix for the system
(\ref{general_model}), viz., $\,Z_6\,$ and $\,\Delta\equiv4Z_3Z_6-Z_5^2\,$
(see Ref.\cite{our_PLB94}
for details).

The preferred way of re-grouping is obvious from the following chain:
\begin{eqnarray}
S&=& \int\!d^2x\,\sqrt{-g}\,\Biggl\{-Z_6\Psi_1^2-2Z_6\Psi_1\left[R+{Z_4\over2
          Z_6}(\partial\Phi)^2+{Z_5\over2Z_6}\square\Phi\right] \qquad\return
  && +\left[Z_1-{Z_4^2\over4Z_6}\right](\partial\Phi)^4+\left[Z_2-{Z_4Z_5
          \over2Z_6}\right](\square\Phi)(\partial\Phi)^2     \qquad\return
  && +\left[Z_3-{Z_5^2\over4Z_6}\right](\square\Phi)^2+\ldots\Biggr\}
                                                                    \\[10pt]
 &=& -\int\!d^2x\,\sqrt{-g}\,\Biggl\{Z_6\Psi_1^2+2Z_6\Psi_1\left[R+{Z_4\over2
          Z_6}(\partial\Phi)^2+{Z_5\over2Z_6}\square\Phi\right] \qquad\return
  && +{\Delta\over4Z_6}\Psi_2^2+{\Delta\over2Z_6}\Psi_2\left[\square\Phi+
          {2Z_2Z_6-Z_4Z_5\over\Delta}(\partial\Phi)^2\right]   \qquad\return
  && +{\Xi\over4Z_6\Delta}\Psi_3^2+{\Xi\over2Z_6\Delta}\Psi_3(\partial\Phi)^2
           -\ldots\Biggr\} \ .
\label{map}
\end{eqnarray}
Here we have introduced the function
\begin{equation}
\Xi(\Phi)=\left(4Z_1Z_6-Z_4^2\right)\Delta-\left(2Z_2Z_6-Z_4Z_5\right)^2 \ .
\end{equation}
It is a trivial matter to verify that $\,\Xi\to\Xi\exp(-4\sigma)\,$ under
the conformal transformations. After a few integrations by parts one arrives
at the structure (\ref{sigma_model}). As anticipated, the field
$\,\Psi_3\,$ does not have a full-fledged tree propagator and its derivatives
contribute exclusively to vertices.

As an aside, let us note that the different versions of the general model
(\ref{general_model}) can be classified into several sets, depending on
whether the functions $\,Z_6(\Phi)\,$, $\,\Delta(\Phi)\,$, or $\,\Xi(\Phi)\,$
are zeroes or not. No conformal metric rescaling or local dilaton field
re-definition may cause either of them vanish.

Equation (\ref{map}) defines a map of the general higher-derivative
scalar-tensor gravity (\ref{general_model}) to a four-dilaton version of
gravity (\ref{sigma_model}) with strong sigma-model motives
\cite{dilaton_gravity_as_string,2-dilaton_models}.

\section{One-loop effective action in dilaton gravity with $R^2$-term}

In this section, we will show that equivalent low derivative model
maybe very useful to do the quantum calculations in the situation
when the calculations in terms of original model are extremely difficult
to do.In particulary,
                 we complete the study of one-loop divergences in the
scalar-tensor higher-derivative gravity (\ref{general_model}) initiated in
\cite{our_PLB94}.The degenerate case of this theory
will be considered as an example.
Consider the following action \cite{degenerate_R2-gravity}:
\begin{eqnarray}
S=\int\!d^2x\,\sqrt{-g}\,\left[{1\over2}Z(\Phi)\partial^\mu\Phi\partial_\mu
   \Phi+C(\Phi)R+V(\Phi)+\omega(\Phi)R^2 \right.   \qquad\qquad \cr
\left. -{1\over2}f(\Phi)\partial^\mu\chi_j\partial_\mu\chi_j\right] \ ,
\label{basic_action_in_Delta=0_class}
\end{eqnarray}
where we have added $N$ real conformal scalar fields $\chi_j\,$. All the
functions of the dilaton, $\Phi$, are assumed to be analytic.  Note that, in
principle, an arbitrary dilaton-curvature coupling, $\,C(\Phi)R\,$, can be
reduced to the linear one, $\,\Phi R\,$, by an appropriate redefinition
of the $\Phi$-field. However, we do not take this option since it would not
facilitate our analysis much.

Basically, the model (\ref{basic_action_in_Delta=0_class}) belongs to a
different class from that discussed in the previous section and in
Ref.\cite{our_PLB94}
since it has $\,\Xi,\Delta=0\,$. An independent calculation is needed in
order to find the divergent structure of the model
(\ref{basic_action_in_Delta=0_class}) which is very difficult to do.
We will show how this calculation maybe done in terms of low
derivative model what is used to restore the one-loop effective
action in original theory.

Since the above action contains higher derivatives, which is difficult to
deal with, we prefer to lower the order of derivatives in the metric sectors
by introducing an auxiliary scalar
\begin{equation}
\Psi=2\omega R \ .
\label{constraint}
\end{equation}
The action under consideration becomes:
\begin{eqnarray}
S=\int\!d^2x\,\sqrt{-g}\,\left\{{1\over2}Zg^{\mu\nu}\partial_\mu\Phi
   \partial_\nu\Phi+\bigl[C+\Psi\bigr]R+V-{1\over4\omega}\Psi^2 \right.
                                                          \qquad\qquad \cr
\left. -{1\over2}fg^{\mu\nu}\partial_\mu\chi_j\partial_\nu\chi_j\right\} \ .
\label{lowered_order_action}
\end{eqnarray}

Within the background field formulation, we use the conformal parametrization
of the metric fluctuations,
\begin{equation}
g_{\mu\nu}\to[\exp\sigma]\,g_{\mu\nu} \ ,
\label{conformal_parametrization}
\end{equation}
and the linear splitting for the scalars
\begin{equation}
\Phi\to\Phi+\varphi \ , \qquad  \Psi\to\Psi+\psi \ , \qquad
\chi_k\to\chi_k+\eta_k \ .
\end{equation}
The ghost contribution which corresponds to
(\ref{conformal_parametrization}), yields Polyakov's term for the conformal
anomaly,
\begin{equation}
{1\over4\pi\epsilon}\int\!d^2x\,\sqrt{-g}\,{13\over3}R+\mbox{finite terms,}
\qquad  \epsilon\to+0 \ .
\label{conformal_ghosts}
\end{equation}

Expanding (\ref{lowered_order_action}) in powers of the quantum
fluctuations, $\{\varphi;\,\psi;\,\sigma;\,\eta_k\}$, one obtains:
\begin{equation}
{S^{(2)},}_{ij}=-\widehat{K}_{ij}\square+\widehat{L}^\lambda_{ij}\nabla_
                  \lambda+\widehat{M}_{ij} \ ,
\end{equation}
where
\begin{equation}
\widehat{K}=\pmatrix{Z & 0 & C' & 0\cr
                     0 & 0 & 1 & 0 \cr
                     C' & 1 & 0 & 0 \cr
                     0 & 0 & 0 & -f\delta_{jk} \cr} \ ,
\end{equation}
and
\begin{eqnarray}
&&\widehat{L}^\lambda_{\varphi\varphi}=-Z'(\partial^\lambda\Phi) \ ,
\qquad
\widehat{L}^\lambda_{\sigma\varphi}=-2C''(\partial^\lambda\Phi) \ , \return
&&\widehat{L}^\lambda_{\varphi\eta_k}=-\widehat{L}^\lambda_{\eta_k\varphi}=
                                       -f'(\partial^\lambda\chi_k) \ ,
\qquad
\widehat{L}^\lambda_{\eta_j\eta_k}=f'(\partial^\lambda\Phi)\delta_{jk} \ ,
\cr\noalign{\vspace{-4pt}} && \\[-2pt]
&&\widehat{M}_{\varphi\varphi}=C''R-Z'(\square\Phi)-{1\over2}Z''(\partial
                                \Phi)^2-{1\over2}f''(\partial\chi_k)^2+V''
                                -\left({1\over4\omega}\right)''\Psi^2 \ ,
                                                                    \return
&&\widehat{M}_{\varphi\psi}=\widehat{M}_{\psi\varphi}={\omega'\over2\omega^2}
                                                          \Psi \ ,   \qquad
\widehat{M}_{\psi\psi}=-{1\over2\omega} \ ,     \qquad
\widehat{M}_{\psi\sigma}=\widehat{M}_{\sigma\psi}=-{1\over2\omega}\Psi \ ;
\nonumber
\end{eqnarray}
other relevant matrix elements are zeroes.

Employing the Schwinger-DeWitt technique%
\footnote{The technical details of such a calculation in a specifically
two-dimensional setting may be found in
\cite{1-loop_treatment_of_dilaton_gravity}.}
and adding (\ref{conformal_ghosts}), we finally get the divergent
contribution to the one-loop effective action, $\Gamma$, for
(\ref{lowered_order_action}):
\begin{eqnarray}
\Gamma_{div}={1\over4\pi\epsilon}\int\!d^2x\,\sqrt{-g}\Biggl\{\left[{23-N
                                                                    \over6}+
               {C''\over Z}\right]R-\left[{1\over\omega}+{C'\omega'\over
               \omega^2Z}\right]\Psi-{(1/\omega)''\over4Z}\Psi^2
                                                                \quad\cr\cr
+{V''\over Z} -{{C'}^2\over2\omega Z}+\left[{N{f'}^2\over4f^2}-{3{Z'}^2
         \over4Z^2}+{Z''\over2Z}\right]g^{\mu\nu}\partial_\mu\Phi\partial_
         \nu\Phi      \quad\cr\cr
-\left[{f''\over2Z}-{{f'}^2\over2fZ}\right]g^{\mu\nu}\partial_\mu\chi_j
         \partial_\nu\chi_j \Biggr\}
\label{Gamma_div}
\end{eqnarray}
(plus non-essential surface terms). The constraint (\ref{constraint})
might be used to eliminate the auxiliary field $\Psi$,
(in a standard background field method way,see Ch.17 in ref.\cite{wes})
which brings about
an unexpected dimension-four operator $\,R^2\,$: the latter is then removed
by the equation of motion $\,\delta S/\delta\Phi=0\,$. (An alternative
approach would be to directly study the sigma-model beta-functions on the
target-space background dictated by the specific embedding
(\ref{constraint}), (\ref{lowered_order_action}).)
That gives the result coinciding with above one on mass shell.

Following the conventional route, one can derive the generalized RG
equations: $\,df/dt=(f''f-{f'}^2)/2fZ\,$, etc. However, these equations are
not very informative because of their cumbersome, non-linear, structure.
Nevertheless, the fixed points of the RG-flow at one loop follow from the
conditions of finiteness%
\footnote{The only subtle point that may show up at higher loops is of the
reparametrization invariance of the cutoff.}
and can be read off Eq.(\ref{Gamma_div}), after the elimination of the
auxiliary field $\Psi$:
\begin{eqnarray}
{11-N\over6}+{C''\over Z}-2{C'\omega'\over\omega Z}=0 \ , \qquad\quad
\left( 1/\omega\right)''=0 \ , \qquad\quad
V''-{{C'}^2\over2\omega}=0 \ , \\[5pt]
{N{f'}^2\over4f^2}-{3{Z'}^2\over4Z^2}+{Z''\over2Z}=0 \ , \qquad\qquad
{{f'}^2\over f}-f''=0 \ .
\end{eqnarray}
These equations are most easily solved for $N=11$, to give
\begin{eqnarray}
f(\Phi)=\alpha_1\exp\left(\alpha_2\Phi/\sqrt{N}\right) \ , \qquad\quad
Z(\Phi)={\alpha_3\exp(\alpha_2\Phi)\over\left[\alpha_4\exp(2\alpha_2\Phi)+1
                                         \right]^2} \ ,              \\[5pt]
\omega(\Phi)=\bigl(\alpha_5\Phi+\alpha_6\bigr)^{-1} \ , \qquad
                                                  \alpha_5\ne0 \ ,   \\[5pt]
C(\Phi)={2\alpha_7\over\alpha_5\Phi+\alpha_6}+\alpha_8 \ , \qquad\quad
V(\Phi)={\alpha_7^2\over\alpha_5\Phi+\alpha_6}+\alpha_9\Phi+\alpha_{10} \ ,
\end{eqnarray}
where $\alpha_1,\ldots,\alpha_{10}$ are arbitrary constants. The above
restrictions are very stringent: notably, the $2D$ successor
\cite{degenerate_R2-gravity}
of the constant-curvature-constrained model%
\footnote{In four dimensions this model offers a way to resolve the unitarity
problem since the negative-norm spin-two states are modded out.}
does not belong to this family and hence cannot be described in terms of
critical strings.

For $\,\alpha_5=0\,$, i.e., $\,\omega\equiv1/\alpha_6=const\,$ another set of
solutions exists:
\begin{equation}
C(\Phi)=2\alpha_7\Phi+\alpha_8 \ , \qquad\quad
V(\Phi)={\alpha_7^2\over\omega}\Phi^2+\alpha_9\Phi+\alpha_{10} \ ,
\end{equation}
while $Z(\Phi)$ and $f(\Phi)$ are the same as before. Yoneya's model
(\ref{Yoneya_model}) corresponds to $\,f,\omega,Z=const\,$ (in which
case $\Phi$ is just another matter scalar) and $\,\alpha_7=\alpha_9=0\,$.
One finds that Yoneya's model coupled to $N$ matter scalars admits the
Virasoro algebra with the effective central charge $\,c=12-N\,$. This value
of $\,c\,$ is in a perfect agreement with the considerations of
Ref.\cite{string_susceptibility}
where the ``string susceptibility'' for (\ref{Yoneya_model}) in the
sigma-model representation was found.

\bigskip
\bigskip

In summary, we have discussed the structure of the higher-derivative
$2D$ dilatonic gravity in terms of a second-derivative model with additional
scalars. Such a consideration suggests the way to map the string theory in
the background of massive modes to the standard sigma-model in a curved
(target) spacetime.

\bigskip
\bigskip

{\bf Acknowledgements}\medskip\par\noindent
We would like to thank A.~Chamseddine, V.~Mukhanov, I.Shapiro
and A.Wipf for useful
discussions. This work was partially supported by DGICYT (Spain), project
SAB93--0024, and RFFR, project 94-02-0324.



\begin{thebibliography}{99}

\bibitem{TASI_lectures}
J.A.~Harvey and A.~Strominger, in: {\sl Recent Developments in Particle
Theory.\,\/}, Proceedings of TASI, Boulder, 1992, eds. J.A.~Harvey and
J.~Polchinski (World Scientific, Singapore, 1993).

\bibitem{our_PLB94}
E.~Elizalde, S.~Naftulin, and S.D.~Odintsov, {\sl Phys. Lett.\/} {\bf B323},
124 (1993).

\bibitem{Buchbinder_etal:book}
I.L.~Buchbinder, S.D.~Odintsov, and I.L.~Shapiro, {\sl Effective action in
quantum gravity.\,\/} (IOP, Bristol and Philadelphia, 1992).

\bibitem{Stelle}
K.~Stelle, {\sl Phys. Rev.\/} {\bf D16}, 953 (1977).

\bibitem{ghost_decoupling_in_HD-gravity}
I.~Antoniadis, C.~Bachas, J.~Ellis, and D.V.~Nanopoulos, {\sl Nucl. Phys.\/}
{\bf B328}, 117 (1989). \\
R.C.~Myers, {\sl Phys. Lett.\/} {\bf B199}, (1987).

\bibitem{Buchbinder_etal:preprint}
I.L.~Buchbinder, E.S.~Fradkin, S.L.~Lyakhovich, and V.D.~Pershin,
{\sl Phys. Lett.\/} {\bf B304}, 239 (1993). \\
I.L.~Buchbinder, V.A.~Krykhtin, and V.D.~Pershin, Preprint TSPI--TH1/94
(1994).

\bibitem{E-expansion_in_gravity}
S.~Weinberg, in: {\sl General Relativity: An Einstein Centenary Survey.\,\/},
eds. S.W.~Hawking and W.~Israel (Cambridge University Press, 1979). \\
R.~Gastmans, R.~Kallosh, and C.~Truffin, {\sl Nucl. Phys.\/} {\bf B133}, 417
(1978). \\
S.M.~Christensen and M.J.~Duff, {\sl Phys. Lett.\/} {\bf B79}, 213 (1978). \\
H.~Kawai and M.~Ninomiya, {\sl Nucl. Phys.\/} {\bf B336}, 115 (1990).
I.~Jack and D.R.T.~Jones, {\sl Nucl.Phys.\/} {\bf B358}, 695(1991).

\bibitem{infrared_quantum_gravity}
I.~Antoniadis and E.~Mottola, {\sl Phys. Rev.\/} {\bf D45}, 2013 (1992). \\
I.~Antoniadis, P.O.~Mazur, and E.~Mottola, {\sl Nucl. Phys.\/} {\bf B388},
627 (1992). \\
S.D.~Odintsov, {\sl Z. Phys.\/} {\bf C54}, 531 (1992).

\bibitem{dilaton_gravity_from_anomaly}
A.M.~Polyakov, {\sl Phys. Lett.\/} {\bf B103}, 207 (1981). \\
R.J.~Reigert, {\sl Phys. Lett.\/} {\bf B134}, 56 (1984).

\bibitem{dilaton_gravity_as_string}
A.H.~Chamseddine, {\sl Phys. Lett.\/} {\bf B256}, 379 (1991); \
{\sl Nucl. Phys.\/} {\bf 368}, 98 (1992). \\
J.~Russo and A.A.~Tseytlin, {\sl Nucl. Phys.\/} {\bf B382}, 259 (1992).

\bibitem{degenerate_R2-gravity}
I.M.~Lichtzier and S.D.~Odintsov, {\sl Mod. Phys. Lett.\/} {\bf A6}, 1953
(1991). \\
T.~Muta and S.D.~Odintsov, {\sl Progr. Theor. Phys.\/} {\bf 90}, 247 (1993).

\bibitem{Yoneya}
T.~Yoneya, {\sl Phys. Lett.\/} {\bf B149}, 111 (1984).

\bibitem{our_IJMPA}
E.~Elizalde, S.~Naftulin, and S.D.~Odintsov, {\sl Int. J. Mod. Phys.\/}
{\bf A9} 933 (1994).

\bibitem{JT-model}
C.~Teitelboim, {\sl Phys. Lett.\/} {\bf B126}, 41 (1983). \\
C.~Teitelboim, in: {\sl Quantum Theory of Gravity.\,\/}, ed. S.~Christensen
(Hilger, Bristol, 1984) p.327;  \
R.~Jackiw, {\sl ibidem\/}, p.403.

\bibitem{2-dilaton_models}
R.B.~Mann, {\bf Phys. Rev.\/} {\bf D47}, 4438 (1993); \
{\sl Nucl. Phys.\/} {\bf B418}, 231 (1994).

\bibitem{1-loop_treatment_of_dilaton_gravity}
S.D.~Odintsov and I.L.~Shapiro, {\sl Phys.Lett.\/} {\bf B263}, 183 (1991); \
{\sl Int. J. Mod. Phys.\/} {\bf D1}, 571 (1993). \\
R.~Kantowski and C.~Marzban, {\sl Phys. Rev.\/} {\bf D46}, 5449 (1992).

\bibitem{string_susceptibility}
H.~Kawai and R.~Nakayama, {\sl Phys. Lett.\/} {\bf B306}, 224 (1993). \\
J.~Nishimura, S.~Tamura, and A.~Tsuchiya, {\sl Mod. Phys. Lett.\/} {\bf A9},
3565 (1994).

\bibitem{sha}
I.L.~Shapiro, {\sl hep-th\/} 9501121.

\bibitem{vol}
I.~Volovich, {\sl Mod.Phys.Lett.\/} {\bf A8}, 1827 (1993).\\
W.~Kummer and D.~Schwarz, {\sl Nucl.Phys.\/}  {\bf B382}, 171 (1992).

\bibitem{wes}
P.~West, {\sl Introduction to supersymmetry and supergravity\,\/},
World Scientific, (1986).

\bibitem{boul}
D.V.~Boulatov and V.A.~Kazakov, {\sl Phys.Lett.\/} {\bf B184}, 247
(1987);
J.~Ambjorn, J.~Jurkiewicz  and C.~Kristiansen, {\sl Nucl.Phys.\/}
{\bf B393}, 601 (1993);
 S.~Ichinose, N.~Tsuda and T.~Yukawa, KEK preprint (1995).

\end{thebibliography}
\end{document}